\magnification\magstephalf
\overfullrule 0pt
\input epsf.tex

\font\rfont=cmr10 at 10 true pt
\def\ref#1{$^{\hbox{\rfont {[#1]}}}$}


\font\twelvebf=cmbx12

\font\tenrm=cmr10 scaled\magstep0


\def\pmb#1{\setbox0=\hbox{#1}
 \kern.05em\copy0\kern-\wd0 \kern-.025em\raise.0433em\box0 }

\def \half {{\scriptstyle {1 \over 2}}}

 %


\def\boxit#1{\vbox{\hrule\hbox{\vrule\kern1pt\vbox
{\kern1pt#1\kern1pt}\kern1pt\vrule}\hrule}}

\def\h{\hfill\break}
\parskip=6pt
\parindent=0pt
\hsize=17truecm\hoffset=-5truemm
\voffset=-1truecm\vsize=24.5truecm
\def\footnoterule{\kern-3pt
\hrule width 17truecm \kern 2.6pt}


\catcode`\@=11 

\def\nolabels{\def\wrlabeL##1{}\def\eqlabeL##1{}\def\reflabeL##1{}}
\def\writelabels{\def\wrlabeL##1{\leavevmode\vadjust{\rlap{\smash%
{\line{{\escapechar=` \hfill\rlap{\sevenrm\hskip.03in\string##1}}}}}}}%
\def\eqlabeL##1{{\escapechar-1\rlap{\sevenrm\hskip.05in\string##1}}}%
\def\reflabeL##1{\noexpand\llap{\noexpand\sevenrm\string\string\string##1}}}
\nolabels
\global\newcount\refno \global\refno=1
\newwrite\rfile
\def\defref{$^{{\hbox{\rfont [\the\refno]}}}$\nref}
\def\nref#1{\xdef#1{\the\refno}\writedef{#1\leftbracket#1}%
\ifnum\refno=1\immediate\openout\rfile=refs.tmp\fi
\global\advance\refno by1\chardef\wfile=\rfile\immediate
\write\rfile{\noexpand\item{#1\ }\reflabeL{#1\hskip.31in}\pctsign}\findarg}
\def\findarg#1#{\begingroup\obeylines\newlinechar=`\^^M\pass@rg}
{\obeylines\gdef\pass@rg#1{\writ@line\relax #1^^M\hbox{}^^M}%
\gdef\writ@line#1^^M{\expandafter\toks0\expandafter{\striprel@x #1}%
\edef\next{\the\toks0}\ifx\next\em@rk\let\next=\endgroup\else\ifx\next\empty%
\else\immediate\write\wfile{\the\toks0}\fi\let\next=\writ@line\fi\next\relax}}
\def\striprel@x#1{} \def\em@rk{\hbox{}} 
\def\lref{\begingroup\obeylines\lr@f}
\def\lr@f#1#2{\gdef#1{\defref#1{#2}}\endgroup\unskip}
\newwrite\lfile
{\escapechar-1\xdef\pctsign{\string\%}\xdef\leftbracket{\string\{}
\xdef\rightbracket{\string\}}}

\def\writestop{\def\writestoppt{\immediate\write\lfile{\string\p
ageno%
\the\pageno\string\startrefs\leftbracket\the\refno\rightbracket%
\string\def\string\secsym\leftbracket\secsym\rightbracket%
\string\secno\the\secno\string\meqno\the\meqno}\immediate\closeout\lfile}}
\def\writestoppt{}\def\writedef#1{}
\catcode`\@=12 

\input amssym.def
\input amssym.tex
\centerline{\twelvebf INTRODUCTION TO}
\vskip 5truemm
\centerline{\twelvebf THERMAL FIELD THEORY\footnote{${}^{*}$}{{\tenrm 
Lectures given in
February 1997 at the IX Jorge Andr\'e Swieca Summer School, Campos do
Jord\~ao, Brazil}}}
\bigskip
\centerline{P V Landshoff}
\centerline{DAMTP, University of Cambridge}
\centerline{pvl@damtp.cam.ac.uk}
\vskip 30pt
\leftline{{\bf Abstract}}
\medbreak
Within the next few years experiments at RHIC and the LHC will seek
to create in the laboratory a quark-gluon plasma, the phase of matter in
which the Universe was initially created. It is believed that the
plasma will survive long enough to reach thermal equilibrium.
I give an introduction to the formalism of thermal field theory, the
combination of statistical mechanics and quantum field theory needed to
describe the plasma in thermal equilibrium,  in a
way that tries to keep close to the physics it describes.

\vskip 30pt
{\bf Introduction}
\medbreak
Thermal field theory is a combination of quantum field theory and
statistical mechanics.  This means that it is both difficult and
interesting.  The reason that we study it is that we want to describe
the quark-gluon plasma, the phase that matter is believed to take
above some critical temperature $T_c$.  Lattice calculations
suggest\defref\lattice{
For a recent review, see A Ukawa, hep-lat/9612011
} that $T_c$ is about $100\; {\rm MeV}$, or $10^{12}$K.
In the plasma phase the quarks and gluons are deconfined; they can
move rather freely through the whole plasma.  This is the phase in
which the universe was created at the big bang, and before the end of
the century experiments at the new collider RHIC will try to re-create
it in the laboratory, by making gold nuclei collide together head-on
and dump their kinetic energy into a small volume. Similar experiments,
at much higher energy, are planned later  for the LHC at CERN.

There is an obvious question: if a plasma is indeed produced, how will
we know it? As yet there is no simple answer. There are estimates\defref\eq{
K J Eskola nd X N Wang, Physical Review D49 (1994) 1284
},
necessarily based on very crude non-equilibrium theory, that suggest
that the plasma will survive for a time long enough that it reaches
thermal equilibrium before it eventually decays back into ordinary matter.
So far, it is only equilibrium thermal field theory that is well formulated,
and my lectures concentrate on this.  For more
information, there is a book that was published last year\defref\lebellac{
M Le Bellac, {\sl Thermal field theory}, Cambridge University Press (1996)
} and
is already the standard text.  As I want my description to stay as
close as possible to physics I will develop the theory using operators
rather than path integrals, and mostly I will use the so-called real-time
formalism.

Because in relativistic theory particles are continually being created
and destroyed, it is appropriate to use the grand partition function

$$
Z = \sum_{i} \langle i | e^{-\beta (H - \mu N)} |i \rangle \eqno{(1)}
$$

Here $\beta$ is the inverse temperature, $\beta = 1/k_B T$, and
usually we use units in which Boltzmann's constant $k_B = 1$.  The
system's Hamiltonian is $H$ and $N$ is some conserved quantum number,
such as baryon number, with $\mu$ the corresponding chemical
potential.  There may be several conserved quantum mumbers, in which
case $\mu N$ is replaced with $\displaystyle{\sum_{\alpha}} \mu_{\alpha} N_{\alpha}$.

The states $|i\rangle$ are a complete orthonormal set of physical states of
the system.  In scalar field theory all states are physical and so

$$
Z = {\rm tr} \; e^{-\beta (H- \mu N)}\eqno{(2)}
$$

which is invariant under changes in the choice of orthonormal basis of
states.  In the case of gauge theories there are unphysical states,
for example longitudinally-polarised photons or gluons, which must be
excluded from the summation in (1).  So then

$$
Z = {\rm tr} \;{\Bbb P}\; e^{-\beta (H-\mu N)} \eqno{(3)}
$$

where ${\Bbb P}$ is a projection operator onto physical states.  The
presence of ${\Bbb P}$ can make things more complicated, and so to begin
with I will consider scalar filed theory, where it is not needed.

All the macroscopic properties of the system in thermal equilibrium
may be calculated from $Z$.  In particular, for a system that is so
large that its surface energy is negligible compared with its volume
energy, the equation of state is

$$
PV = T \; \log\; Z \eqno{(4)}
$$

Also, the ``thermal average'' of an observable corresponding to an
operator $Q$ is

$$
<Q> = Z^{-1}\; {\rm tr}\; Q e^{-\beta (H-\mu N)}  \eqno{(5)}
$$

Notice that, throughout, all operators are familiar zero-temperature
ones.  The temperature enters only in the exponential, which
characterises the particular ensemble of states used to calculate
expectation values of the operators.
\bigskip
{\bf Noninteracting scalar bosons}

For many systems of bosons there is no conserved quantum number $N$;
for example, in the case of a heat bath of photons there is no
constraint on their total number.  Then in the scalar-field-theory
case $Z$ is just ${\rm tr}\; e^{-\beta H}$. 

In the absence of interactions, the energies of the separate particles
are good quantum numbers.
To begin with, quantise the system in a finite volume $V$, so that the
single-boson energies $\epsilon_r$ are discrete.  The states $|i\rangle$ of
the system are labelled by the single-particle occupation numbers
$n_r$, and the eigenvalues of the noninteracting Hamiltonian $H_o$ are

$$
n_1 \epsilon_1 + n_2 \epsilon_2 + n_3 \epsilon_3 + \dots
$$

So the noninteracting grand partition function is 

$$
\eqalign{
Z_0 &= \sum_{\{n_r\}}  e^{\beta (n_1 \epsilon_1 + n_2 \epsilon_2 +\dots )} \cr
&=  \prod_r \Big (\sum _{n_r}e^{-\beta n_r e_r}\Big ) = \prod_r {1\over
1-e^{-\beta \epsilon_r}}\cr} \eqno (6a) 
$$

and so

$$
\log Z_0 = - \sum_r \log \;(1 -e^{-\beta\epsilon_r}) \eqno (6b) 
$$

In the continuum limit

$$
\sum_r \rightarrow V \int {d^{3}k \over (2\pi)^3} \eqno (7) 
$$

and so the noninteracting equation of state is

$$
P = {T\over V} \log Z_0 = -T\int {d^3 k \over (2\pi )^3} \log (1
- e^{-\beta k_0)} \eqno (8)
$$

where $k_0 = \sqrt{{\bf k}^2 + m^2}$.  (If the bosons have non zero
spin, there is an additional factor $g_s$ corresponding to the spin
degeneracy of each single-boson state.)

In the continuum limit we usually work with fields

$$
\phi (x) = \int {d^{3}k\over (2\pi)^3}\; {1\over 2k_0} a({\bf k})e^{-ik\cdot
x} + h.c. \eqno (9a)
$$

with

$$
[a ({\bf k}), a^{\dag} ({\bf k'})] = (2\pi)^3 2k_0 \delta^{(3)} ({\bf
k}-{\bf k'})\eqno (10a)
$$  

In the discrete case, we usually define

$$
[ a_r, a^{\dag}_s ] = \delta_{rs} \eqno (10b)
$$

If we sum this over $r$, the result is $1$.  But if we apply $V \int
d^3 k/(2\pi)^3$ to $(10a)$, the result is rather $2k^0 V$.  That is
(10a) and (10b) have definitions of the operators $a$ differing by a
factor $\sqrt{2k_0 V}$.  We correct for this by defining the field in
the discrete case to be

$$
\phi (x) = \sum_r {1\over \sqrt{2\epsilon_r V}} a_r e^{-i\epsilon _
r{t}} e^{i{\bf k}_r \cdot {\bf x}} + h.c. \eqno (9b)
$$

We can now calculate the thermal average

$$
\langle T \phi (x) \phi (0) \rangle_0 
= Z^{-1}_{0} \sum_i \langle i| e^{-\beta H_0} T
\phi (x) \phi (0) |i\rangle \eqno {(11)}
$$

Observe first that

$$
T \phi (x) \phi (0) = \langle0| T \phi(x) \phi (0) |0\rangle + : \phi (x) \phi
(0): \eqno {(12)}
$$

where : : denotes the usual normal product. The first term contributes
to (11) just the usual zero-temperature Feynman propagator.  To
evaluate the contribution from the second, use the discrete case (9b)
and so obtain double sums $\displaystyle{\sum_{r,s}}$ of terms $a^{\dag}_r a_s
, a_r a_s,
a^{\dag}_r a^{\dag}_s$.  When we take the necessary expectation values, only
the first survives, and then only for $r=s$.  We may replace $a^{\dag}_r
a_r$ with $n_r$, and use

$$
\left ({1\over 1-e^{-\beta\epsilon}}\right )^{-1}
\; \sum_n n \; e^{-n\beta\epsilon} =
f(\epsilon)\eqno (13a)
$$

where $f$ is the Bose distribution

$$
f(\epsilon ) = {1\over e^{\beta\epsilon} -1} \eqno (13b)
$$

So we find that

$$
\langle: \phi (x) \phi (0):\rangle_0 = {1\over V} \sum_r {1\over 2\epsilon_r}
f(\epsilon_r) e^{i\epsilon_r t - i{\bf k}r \cdot {\bf x}} + h.c. \eqno
(14)
$$

Going to the continuum limit, we have

$$
\eqalign{
\langle T \phi (x) \phi (0) \rangle_0 &= \int {d^4 k\over (2\pi)^4}\;
e^{-ik\cdot x} D^0_T (k)\cr
D^0_T (k) = {i\over k^2 -m^2 + i\epsilon} &+ 2\pi \delta (k^2 - m^2)\,
n(k^0)\cr} \eqno (15) 
$$
where $n(k^0) = f(|k^{0}|)$.  The second term is the contribution from
the heat bath; it contains the $\delta$-function because so far the
heat-bath particles do not interact and so they are on shell.
\bigskip
{\bf Perturbation theory}

Suppose now that we introduce an interaction and again calculate
$\langle T \phi (x) \phi (0) \rangle$.  Now $\phi (x)$ is the
interacting Heisenberg-picture field: it is the familiar operator of
zero-temperature field theory.  We can develop a perturbation theory
along the same lines as at zero temperature, by introducing the
interaction picture that coincides with the Heisenberg picture at some
time $t_0$:

$$
\eqalign{
\phi_I (t, {\bf x}) &= \Lambda (t)\phi (t, {\bf x}) \Lambda^{-1} (t) \cr
\Lambda(t) &= e^{i (t - t_0)H_{0I} } e^{-i(t-t_0)H } \cr} \eqno (16)
$$
where $H_{0I}$ is the free-field Hamiltonian in the interaction
picture.  As usual

$$
\eqalign{
\Lambda (t_1) \Lambda^{-1} (t_2) &= U(t_1, t_2) \cr
&= T \exp \left ( -i \int^{t_1}_{t_2} dt\; H^{{\rm INT}}_I (t) \right )
\cr} \eqno (17)
$$

We need $U(t_1, t_2)$ for complex $t_1$ and $t_2$, so that we
integrate $t$ along some contour running from $t_2$ to $t_1$ in the
complex plane and generalise the time ordering $T$ to an ordering
$T_c$ along $C$: The operator whose argument is nearest to $t_1$ along
the contour comes first.

Now, from the definition of $U$,

$$
e^{-\beta H} = e^{-\beta H_{0I}} \; U(t_0 -i\beta, t_0)
$$ 

so that

$$
\eqalign{
Z^{-1} {\rm tr} \; e^{-\beta H} \phi (x) \phi (0) &= 
 Z^{-1} {\rm tr}\; e^{-\beta H_{0I}} U(t_0 -i\beta , x^{0})\,\phi_{I}(x) 
U(x^0, 0)\, \phi_I (0)\, U (0, t_0) \cr
&= Z_0 Z^{-1}\big\langle U(t_0 - i\beta , x^0)\, \phi_I (x)\, U (x^{0}, 0)
\phi _I(0)\, U(0, t_{0}) \big\rangle_0 \cr}\eqno (18) 
$$
where I have used $U(t_1,t_2)U(t_2,t_3)=U(t_1,t_3)$.

Compare with what one has at zero temperature:

$$
\langle 0 | \phi (x) \phi (0) |0\rangle = \langle 0 | U (\infty,
x^0) \phi_I (x) U (x^{0},0 ) \phi_I (0) U(0,-\infty ) |0\rangle
$$

In (18) we have a non-interacting thermal average instead of a vacuum
expectation value, $(t_0 -i\beta )$ instead of $\infty$, and $t_0$
instead of $-\infty$.  In fact there is a close similarity between
thermal perturbation theory and the usual zero-temperature Feynman
perturbation theory.  The main difference is that instead of the
internal lines in the ordinary Feynman graphs representing vacuum expectation
values, in the thermal graphs 
they are non-interacting thermal averages.  In order to derive
this, one needs to establish Wick's theorem.  It is a remarkable fact
that indeed, for example,

$$
\langle T_c \phi_I (x_1) \phi_I (x_2) \phi_I (x_3) \phi_I (x_4)
\rangle_{0} = \sum \langle T\phi_I \phi_I \rangle_0\, \langle T
\phi_I \phi_I \rangle_0 \eqno (19)
$$

where the sum is over the possible pairings of the fields.  For almost
every ensemble other than one in thermal equilibrium there would be
correction terms to (19).

One needs to choose a value for $t_0$.  A common choice is $t_0 = 0$,
with the contour $C$ for the $t$ integrations running from $0$ to
$-i\beta$ along the imaginary axis.  This is the imaginary-time
formalism.  Alternatively, $t_0 \rightarrow -\infty$,  which with suitable
contour choice gives the real-time formalism.
\bigskip
{\bf Real-time formalism}

One is interested in equilibrium properties of the plasma at finite
times.  Presumably these are independent of how it reached thermal
equilibrium.  So, as in familiar scattering theory, we are free to
imagine that the interaction slowly switches off as we go into the
remote past, and then when we take $t_0\rightarrow -\infty$ the
interaction-picture fields become the usual noninteracting {\sl in}
fields.  The corresponding {\sl in} states are direct products of
non-interacting single-particle states.  We need to choose how the
contour $C$ runs from $-\infty$ to   $(-\infty -i\beta )$, and the
choice that keeps the formalism in the most direct contact with
the physics is the  so-called Keldysh
one: along the real axis from $-\infty$ to $\infty$, back to
$-\infty$, then straight down to $(-\infty -i\beta )$.
$$
\matrix{\hbox{\epsfxsize=80truemm  \epsfbox{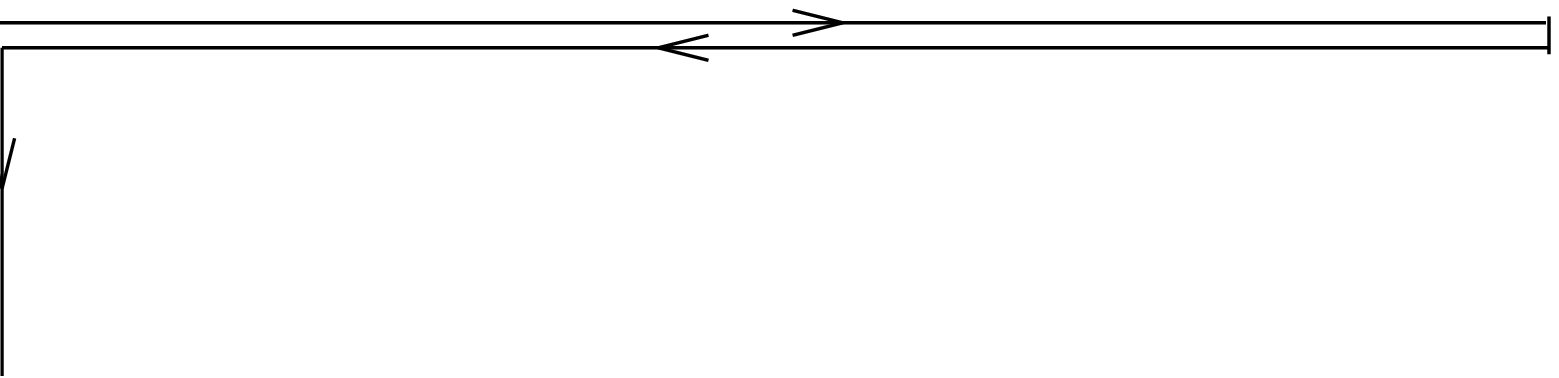}}\cr}
$$

For most applications (though not all\defref\evans{
T S Evans and A C Pearson, Physical Review D52 (1995) 4652
}) it turns out that the
vertical part of the contour does not contribute.  Then we may write
the propagator that corresponds to a line of a thermal
graph as a $2\times 2$ matrix:

$$\eqalign{
{\bf D} (x_1, x_2) &= \left [\matrix{\langle T \phi_{{\rm in}} (x_1)
\phi_{{\rm in}} (x_2) \rangle_0 & \langle \phi_{{\rm in}} (x_2)
\phi_{{\rm in}} (x_1) \rangle_0 \cr
\langle\phi_{\rm_{in}} (x_1) \phi_{{\rm in}} (x_2) \rangle_0 &
\langle\bar{T} \phi_{{\rm in}}(x_1)\phi_{{\rm in}} (x_2) \rangle_0 \cr}\right ]}
\eqno(20)
$$

When both $x_1$ and $x_2$ are on the $-\infty$ to $\infty$ part of
$C$, the ordering $T_c$ is ordinary time ordering $T$; this
corresponds to the element $D_{11}$ of $\bf{D}$.  When both are on
the $\infty$ to $-\infty$ part of $C$, $T_c$ is anti-time-ordering
$\bar{T}$; this corresponds to $D_{22}$.  The off-diagonal elements
correspond to $x_1$ being on one part of $C$ and $x_2$ on the other.
There is translation invariance: the elements of ${\bf D}$ depend only on
the difference between $x_1$ and $x_2$.

I have already shown how to calculate $D_{11}$; the result is given
in (15).  The other elements of $\bf{D}$ may be calculated in the
same way, and its Fourier transform is

$$
{\bf D}(k)=\left[\matrix{i\over {k^2-m^2 + i\epsilon} &2\pi\delta ^-(k^2-m^2)\cr
\delta^+(k^2-m^2)&
{-i\over k^2-m^2-i\epsilon}\cr} \right ]+ 2\pi \delta (k^2 - m^2)\,n(k^0) \ 
\left[\matrix{ 
1 & 1 \cr 1 & 1 \cr}\right]\eqno (21a)
$$

It may also be written in the form

$$
{\bf M} \ \tilde{\bf D} \  {\bf M}\eqno (21b)
$$

with

$$\eqalignno{
\tilde{\bf D} = & i \left[ \matrix{1\over {k^2-m^2 + i\epsilon}& 0\cr
0& {-1\over k^2-m^2-i\epsilon}\cr} \right]\cr
&\cr
{\bf M} = & \sqrt{n(k^0)} \left[\matrix{ e^{{1\over 2} \beta |k^0|} & 
e^{-{1\over 2}\beta k^0} \cr e^{{1\over 2}\beta k^0} & e^{{1\over 
2}\beta |k^0|}\cr}\right]&(21c)} 
$$
For the case of a fermion field, there is a rather similar matrix
propagator, but with the Fermi-Dirac distribution replacing the Bose
distribution.

\bigskip
{\bf Matrix structure}

The elements of the matrix propagator (20) are not independent. For example,
$$
\eqalignno{
D_{21} (x) &= \langle \phi_{{\rm in}} (x) \phi_{{\rm in}} (0)
\rangle_0 = Z_0^{-1} {\rm tr}\; e^{-\beta H_{0{\rm in}}} \phi_{{\rm in}}
(x) \phi_{\rm in} (0) \cr
&= Z_0^{-1} {\rm tr}\; e^{-\beta H_{0{\rm in}}} \phi_{{\rm in}}(0)
e^{\beta H_{0{\rm in}}} \phi_{{\rm in}}(x) e^{{\beta H_0{\rm in}}} \cr
&= Z_0^{-1} {\rm tr}\; e^{-\beta H_{0{\rm in}}} \phi (0) \,\phi (x^0 - i\beta
, {\bf x}) \cr
&= D_{12} (x^0 - i\beta , {\bf x})&(22a)\cr}
$$

Here, I have used a general property of traces, that ${\rm tr}(AB) =
{\rm tr} (BA)$, and the fact that $H_{0{\rm in}}$ is the time-translation
operator for the noninteracting field $\phi_{{\rm in}}$.  
The Fourier transform of (22a) is

$$
D_{12} (k) = e^{\beta k^0} D_{21} ({k}) \eqno (22b)
$$

Also, from their definitions (20), one can see that $D_{11}$ and $D_{22}$
may be expressed in terms of $D_{12}$ and $D_{21}$.  
It is this, together with (22b),
which is responsible for the matrix structure (21).  

Define
a dressed thermal propagator matrix  $ {{\bf D}}' (x_{1}, x_{2})$
analogous to ${\bf D} (x_1, x_2)$ in (20), but with the
interacting Heisenberg field instead of $\phi_{{\rm in}}$.  
For example, $D'_{12} (x_1, x_2)=
\langle \phi (x_2) \phi (x_1)\rangle$.  Then, because $H$ is the
time-translation operator for $\phi$, we can again derive

$$
D'_{12} (k) = e^{\beta k^0} D'_{21} (k) \eqno (22c)
$$

and so deduce that  ${\bf D'}$ has the structure

$$\eqalignno{
{\bf D'} (k) &= {\bf M} \pmatrix{\hat D'(k) & 0 \cr
0 & \hat D^{\prime *}(k)} {\bf M} &(23)\cr}
$$

Define the thermal self-energy matrix ${\bf \Pi}$ by

$$
i{\bf \Pi} = {\bf D'}^{-1} - {\bf D}^{-1} \eqno (24a)
$$

Then ${\bf\Pi}$ has the structure 

$$
i{\bf \Pi} = {\bf M}^{-1} \pmatrix{i\hat\Pi'(k,T) & 0 \cr
0 & [i\hat\Pi^{\prime} (k, T)]^*\cr}{\bf M}^{-1} \eqno (24b)
$$

If we then solve (24a) for ${\bf D'}$, we find

$$
{\bf D}' = {\bf M} \pmatrix {i\over k^2 - m^2 - \Pi & 0 \cr
0 & {-i \over k^2 - m^2 - \Pi^{*}} \cr} {\bf M} \eqno (25)
$$

So it is natural to interpret ${\rm Re}\;\hat \Pi$ as a temperature-%
dependent shift to the mass $m^2$.  $\hat\Pi$ also has an imaginary part, so
the propagation of the field through the heat bath decays with time.

In scalar field theory,
\font\bfit = cmbxti10

$$
{\bfit\Pi}=\matrix{\hbox{\epsfxsize=80truemm  \epsfbox{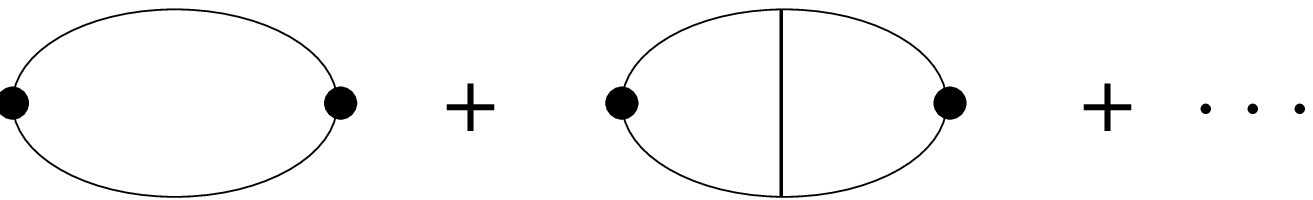}}\cr}
$$

To calculate the contribution to $\Pi_{12}$ from the second term, for
example, one needs

$$
\matrix{\hbox{\epsfxsize=100truemm  \epsfbox{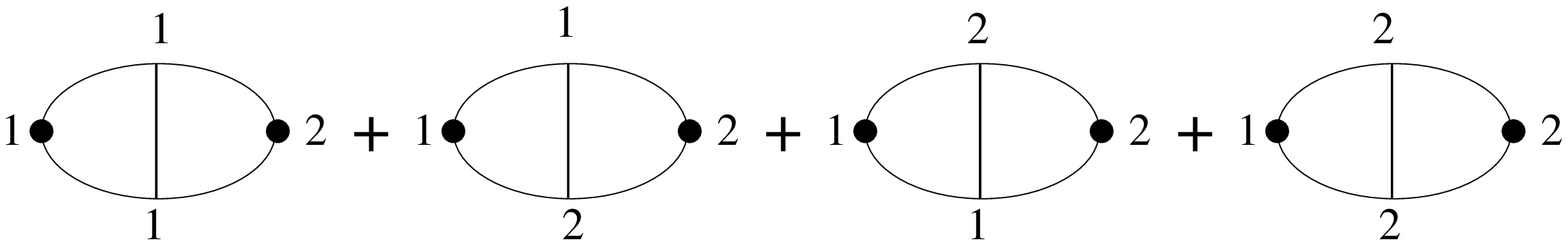}}\cr}
$$

where each line $\matrix{\hbox{\epsfxsize=10truemm  \epsfbox{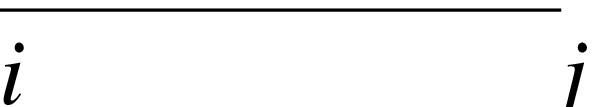}}\cr}$
 represents $D_{ij} (k)$ and each vertex $2$ is
the same as the normal vertex $1$, but opposite in sign.
\bigskip
{\bf Imaginary-time formalism}

The real-time formalism stays close to the physics, but has the calculational
complication that the propagator is a matrix. In the imaginary-time formalism
there is not this complication, though except for a few simple cases
there is  the need to perform an anlytic continuation from imaginary
to real time at the end of the calculation.
In the imaginary-time formalism  
the $t$ integration runs along the imaginary axis, so ordinary
time-ordering is replaced with ordering in imaginary time:

$$
\bar{D} (x_1, x_2) = \theta (-{\rm Im}\; t) D_{21} (x_1, x_2) + \theta
({\rm Im}\; t)
D_{12} (x_{1}, x_{2}) \eqno (26) 
$$

where $t = x_1^0 - x^0_2$.  Because both $x^0_1$ and $x^0_2$ are
integrated from $0$ to $i\beta$, we need $\bar{D} (x_{1},x_{2})$ for
values of ${\rm Im}\; t$ in the range $-\beta$ to $+\beta$.  In this finite
interval it has a Fourier-series expansion.

$$
\bar{D} (t, {\bf x}) = {i\over \beta} \sum^\infty_{n=-\infty} D_n
({\bf x}) \, e^{\omega _n t} \eqno (27)
$$

where $\omega _n = n\pi/\beta$.  However, the relation (22a) implies that
$\bar{D} (t,x) = \bar{D} (t+i\beta, {\bf x})$, so that only even
values of $n$ contribute to the sum.  (In the case of fermions, the
anticommutativity of the fields results in a minus sign appearing in the
corresponding relation (22a), and so then only odd values of $n$
contribute.)

If we apply a $3$-dimensional Fourier transformation to (27) and
invert the Fourier summation over $n$, we find

$$
D_n ({\bf k}) = \int^{i\beta}_{0} dt\; e^{-w_n t} D_{21} (t, {\bf k})
\eqno (28)
$$

which turns out to be just the ordinary Feynman propagator with $k^0 =
i\omega _n$.  So the Feynman rules are just like the zero-temperature ones,
except that the energy-conserving $\delta$-function at each vertex is
replaced with a Konecker delta which imposes conservation of the
discrete energy, and round each loop of a thermal graph

$$
\int {d^4 k\over (2\pi)^4} \rightarrow {i\over\beta} {\sum_n} \int
{d^{3}k\over (2\pi)^3} \eqno (29)
$$

{\bf Gauge theories}

For gauge theories there is the complication that the grand partition
function has to include the projection operator ${\Bbb P}$ onto
physical states: see (3).  There are two formalisms for the resulting
perturbation theory\defref\rebhan{
P V Landshoff and A Rebhan, Nuclear Physics B383 (1992) 607 and B410 (1993) 23 
}: 
{\parindent=8truemm
\item{A} Only the two physical degrees of freedom of the gauge field (the
transverse polarisations) acquire the additional thermal propagator;
the other components of the gauge field, and the ghosts, remain frozen
at zero temperature.  (This is for the bare propagators; self-energy
insertions in the unphysical bare propagators do depend on the
temperature.)
\item{B} All components of the gauge field, and the ghosts, become heated to
temperature $T$.}

In the zero-temperature field theory, the ghosts are introduced in
order to cancel unwanted contributions from the unphysical components
of the gauge field, and the two formalisms lead to the same answers
for calculations of physical quantities for that reason.  Often, using
formalism $A$ makes calculations simpler.  It also makes them stay
closer to the physics.
\bigskip
{\bf Photon or dilepton emission from a plasma}

As an application, consider the emission of a real or virtual photon
of momentum $q$ from a quark-gluon plasma.  This is supposed to be an
important diagnostic test of whether a plasma has been created in an
experiment and reached thermal equilibrium, and a way to measure its
temperature.

Before a photon is emitted, the plasma is described by the density
matrix

$$
\rho = Z^{-1} \sum_i | i\, {\rm in} \rangle\langle i\, {\rm in} | e^{-\beta H} \eqno
(30)
$$

The emission probability is calculated from squared matrix elements of
the Heisenberg-picture electromagnetic current:
$$
W^{\mu\nu} (q) = Z^{-1} \sum_{f} \int d^4 x e^{-iq\cdot x} \langle f
{\rm out} | J^{\mu} (x)\left ( \sum_{i}| i {\rm in}\rangle\langle
 i {\rm in}| e^{-\beta H} \right )J^{\nu} (0) | f {\rm
out\rangle } \eqno (31a)
$$  

where I have introduced also a complete set of out states for the
plasma.  These satisfy the completeness relation

$$
\sum_f | f{\rm out} \rangle\langle f {\rm out} | = 1 
$$

and so
$$
W^{\mu\nu} (q) = Z^{-1} \sum_i \int d^4 x\, e^{iq\cdot x} \langle
i {\rm in}|e^{-\beta H} J^{\nu} (0) J^{\mu} (x) |i {\rm in}\rangle \eqno (31b)
$$

If we introduce a matrix ${\bf G}^{\mu\nu}(q)$ analagous to ${\bf D}'$, but
with thermal averages of products of electromagnetic currents  instead
of fields, $W^{\mu\nu} (q)$ is just $G^{\mu\nu}_{12} (-q)$.  So we draw
thermal graphs where current $-q$ enters at a $1$ vertex and leaves
at a $2$ vertex, and distribute the labels $1$ and $2$ in all possible
ways on the other vertices.  For example,

$$
\matrix{\hbox{\epsfxsize=110truemm  \epsfbox{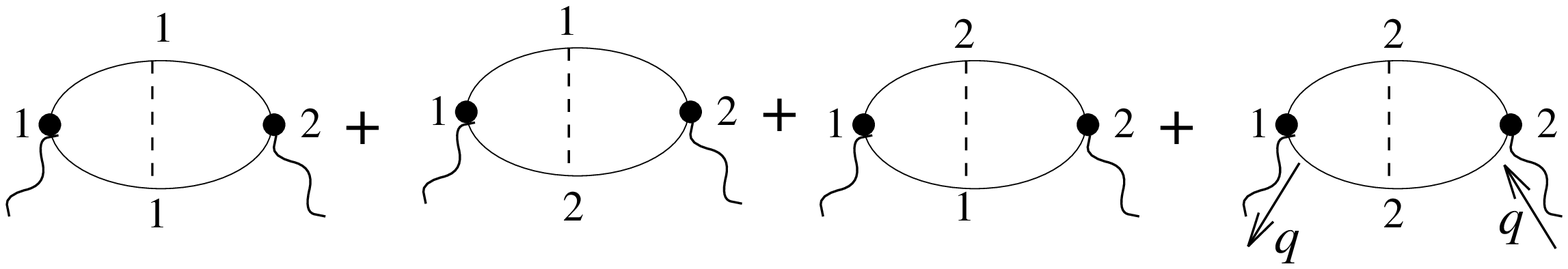}}\cr}
$$

The emission rate is calculated from a matrix element times its complex
conjugate; the vertices labelled 1 correspond to a contribution to
the matrix element and those labelled 2 to its complex conjugate.
The sets of 1-vertices and of 2-vertices are joined by {\sevenrm 12} lines
which, according to (21a) are on shell and represent particles in the heat bath.
The thermal graphs sum together many physical processes.  Consider the
first graph, for example. Its right-hand part represents the contributions
$$
\matrix{\hbox{\epsfxsize=100truemm  \epsfbox{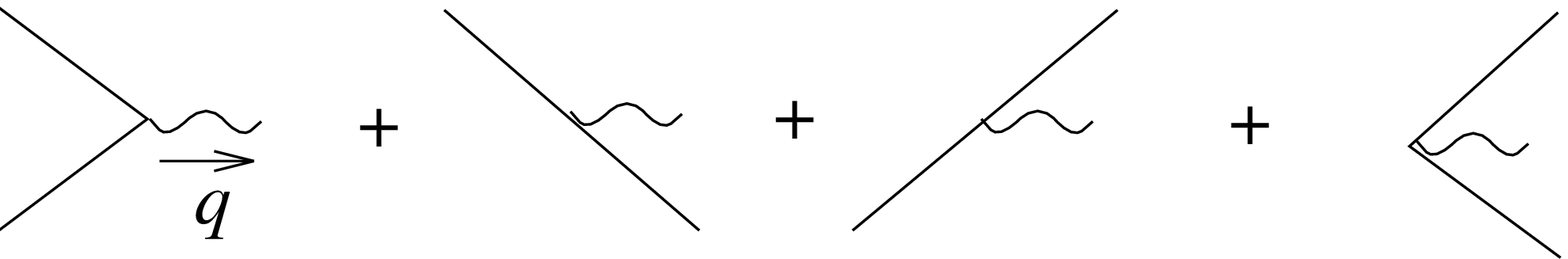}}\cr}
$$

to the amplitude.

In fact, energy-momentum conservation allows only the first one to be
non-zero.  I have not drawn in the other heat-bath particles, but
remember that they are there as spectators.  The other part of the thermal
graph contains only {\sevenrm 11} lines.  If I use only the zero-temperature
part of $D_{11}$ in each, I obtain the amplitude

$$
\matrix{\hbox{\epsfxsize=25truemm  \epsfbox{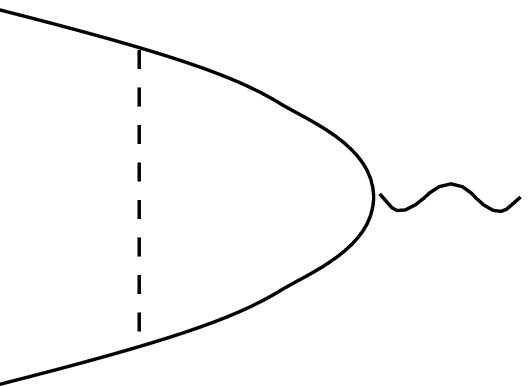}}\cr}
$$
(plus other terms which again vanish for kinematic reasons)
and so part of the thermal graph represents the interference between this and
$\matrix{\hbox{\epsfxsize=8truemm  \epsfbox{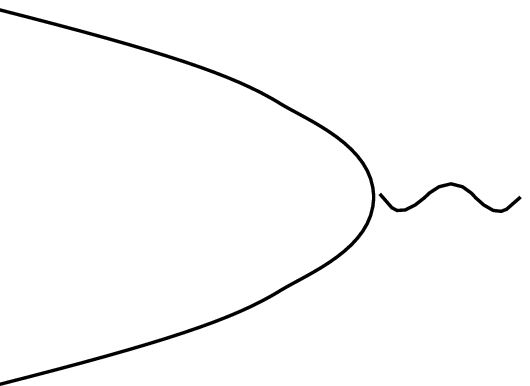}}\cr}$.  If instead I use 
the thermal part $n(k^0)\,2\pi\delta(k^2)$  of the {\sevenrm 11} gluon
propagator, I obtain the amplitudes

$$
\matrix{\hbox{\epsfxsize=70truemm  \epsfbox{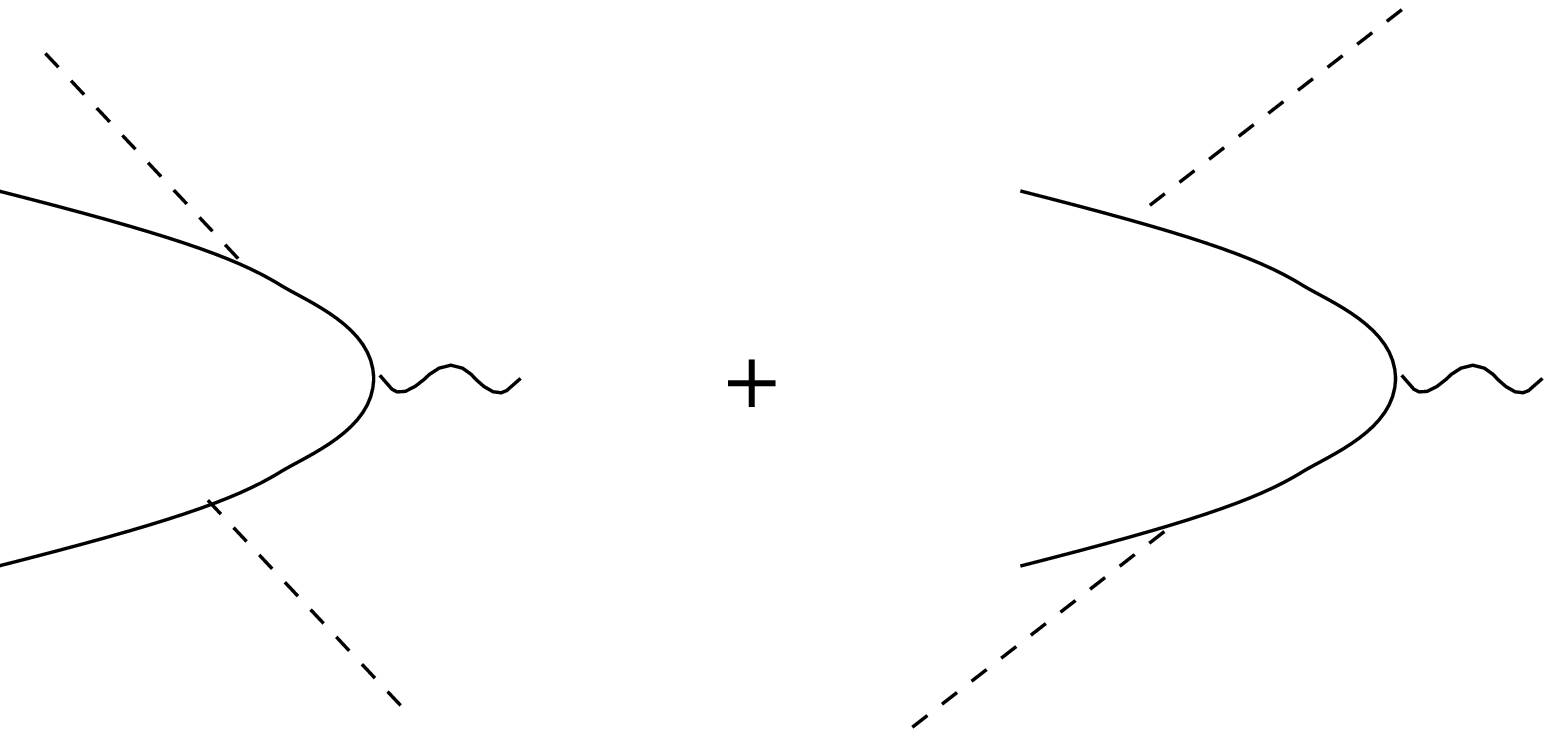}}\cr}
$$

In each case, the incoming and outgoing gluon lines must have the same
momentum $k$, so that these amplitudes again interefere with 
$\matrix{\hbox{\epsfxsize=8truemm  \epsfbox{5a.eps}}\cr}$,
but now with the gluon $k$ being one of the spectator particles in
the heat bath: 

$$
\matrix{\hbox{\epsfxsize=25truemm  \epsfbox{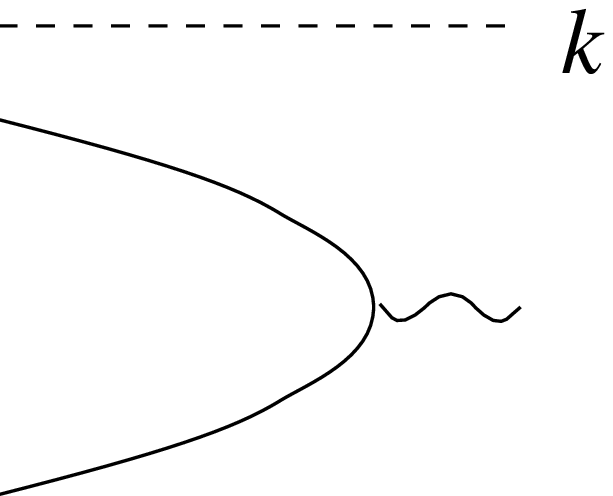}}\cr}
$$

Similarly, I  can identify physical processes that involve the thermal
parts of the {\sevenrm 11} quark propagators.

Even a simple-looking thermal graph corresponds to a large number of
physical processes\defref\niegawa{
N Ashida et al, Physical Review D45 (1992) 2066\h
P V Landshoff, Physics Letters B386 (1996) 291
}, each of which can be rather complicated.  An
example is

$$
\matrix{\hbox{\epsfxsize=45truemm  \epsfbox{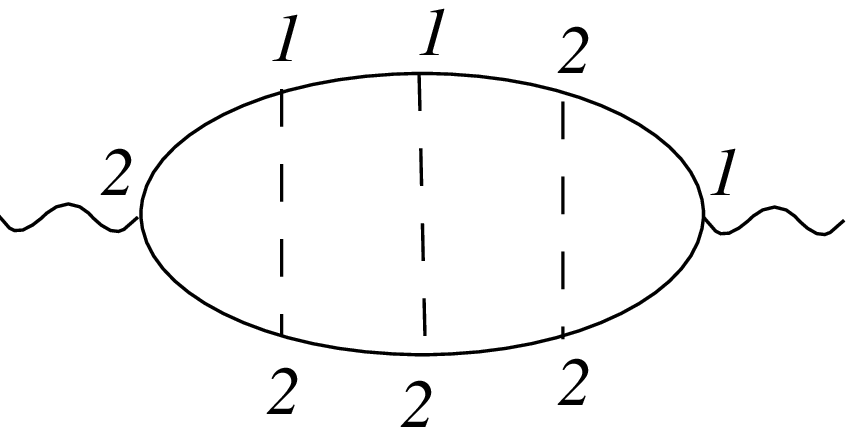}}\cr}
$$

for which just one of the physical processes is the interference between

$$
\matrix{\hbox{\epsfxsize=75truemm  \epsfbox{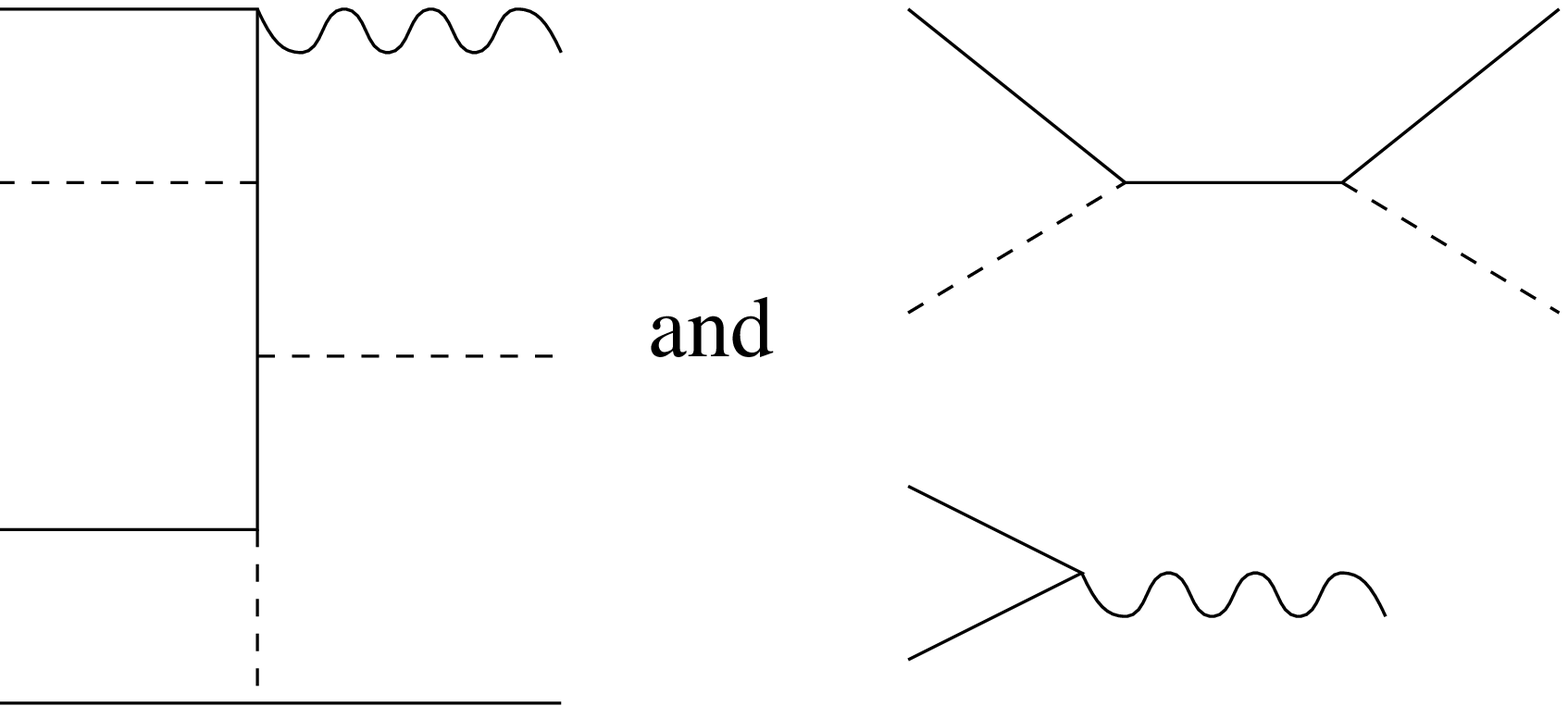}}\cr}
$$  

A disconnected graph occurs because the thermal graph has an
``island'' of a pair of {1}-vertices entirely surrounded by
{2}-vertices
\bigskip
{\bf Infrared divergences}

The infrared divergences of zero-temperature become much worse at
finite temperature: the Bose distribution diverges at zero energy and
causes the usual logarithmic divergences to become power divergences.
We know that the infrared divergences must cancel if the theory is to
make sense, and in practice they always do, but there is no general
theory to show this.  To some extent, the situation can be rescued by
including thermal self-energy insertions in the propagators, so that
they acquire a mass proportional to the temperature.  But, in the case
of photons or gluons, not all the degrees of freedom have a mass,
according to perturbation theory.  If the so-called magnetic mass is
non zero, it is nonperturbative.  

Consider, for example, the effect on
the decay rate $\pi^0 \rightarrow e^+ e^-$ of the microwave
background, which is a heat bath consisting only of photons\defref\jacob{
M Jacob and P V Landshoff, Physics Letters B281 (1992) 114
}.  Let
$\Gamma$ be the decay rate in vacuum.  The heat bath will change it
partly because it gives the electrons an additional
temperature-dependent mass $\delta m^2_e \propto e^2 T^2$.  This causes
a change $\delta\Gamma = \delta m_e^2 \,\partial\Gamma/\partial m^2_e$,
which is associated with thermal graphs of the form

$$
\matrix{\hbox{\epsfxsize=35truemm  \epsfbox{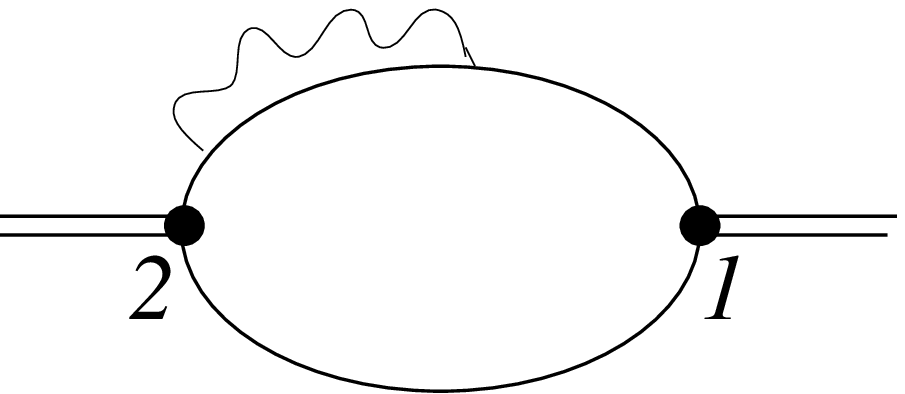}}\cr}
$$

There is also the thermal graph

$$
\matrix{\hbox{\epsfxsize=35truemm  \epsfbox{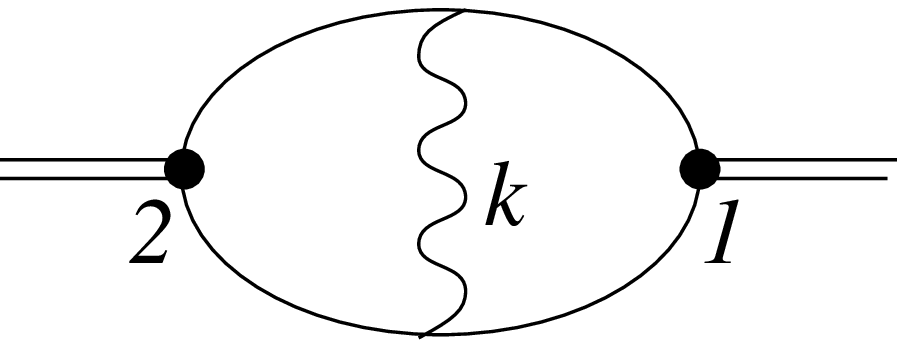}}\cr}
$$

One finds that 

$$
\eqalign{
{\Delta\Gamma\over\Gamma} ={\delta m_e^2\over\Gamma} \,{\partial\Gamma
\over\partial m^2_e}+
& {m_{\pi}\alpha _{EM} \over \pi ^3Q}
 \int d^4 k \,\delta (k^{2}) n(k^0)
\int d^4 p_1 d^4 p_2\, \delta^{(+)} (p_{1}^{2} - m^2_e )
\delta^{(+)} (p^2 - m_e^2 ) \cr
&\qquad\left ( {p_1\over p_1 \cdot k} - {p \over p_2 \cdot k}\right )^2
\big\{ \delta^{(4)} (p_1 + p_2 - P) - \delta^{(4)} (p_1 + p_2 + k - P)
\big\}\cr} \eqno {(32)}
$$
where $Q^2 = ({1\over 4} m^2_{\pi} - m^2_e )$.

In the integral, the first $\delta^{(4)}$-function corresponds to the
contribution from the internal vertices in the thermal graphs
 being both {\sevenrm 1} 
or both {\sevenrm 2}
and the second to {\sevenrm 12} and {\sevenrm 21}.
Each term separately is infrared
divergent, like $\int dk/k^2$, but the divergences  cancel.

In fact there is more cancellation than just that of the infrared
divergences.  For $T \ll Q$ one can expand (32) in powers of $T^2/Q^2$.  
One finds that
the first term, of order $\alpha _{EM} T^2/Q^2$, exactly cancels the
electron-mass-shift contribution $\delta \Gamma$, so that the net
change in the decay rate is of order $\alpha _{EM} T^4$.  The lesson is that,
in thermal field theory, it is of importance to calculate all terms;
gauge theories at finite temperature are rife with cancellations.

Another example where there is an apparent infrared problem is that of
the calculation of the equation of state for the quark-gluon plasma.
For instance, in the purely gluonic case thermal graphs of the form

$$
\matrix{\hbox{\epsfxsize=50truemm  \epsfbox{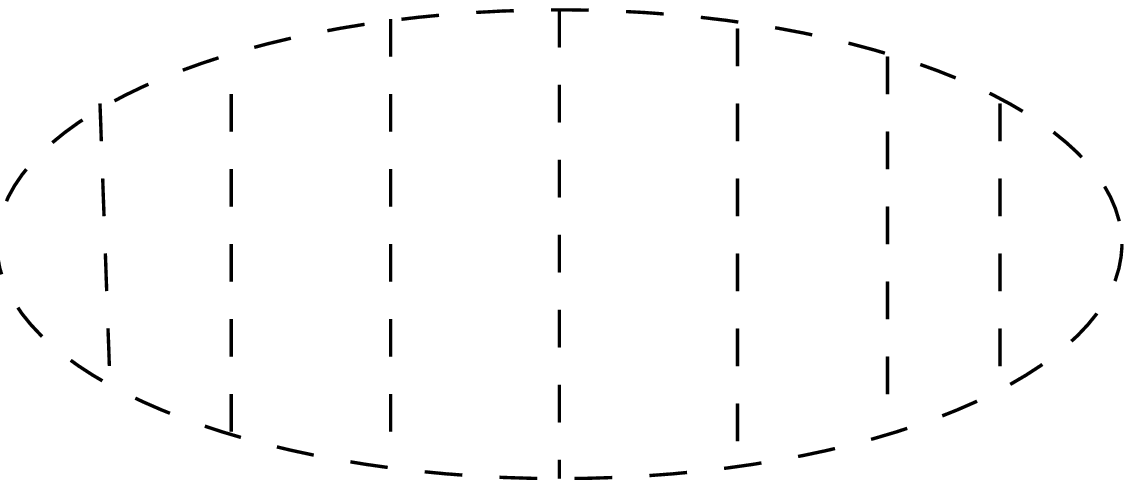}}\cr}
$$

become more and more divergent as more vertical lines are added.  To
see this, use the imaginary time formalism.  The term in the multiple
sum over the energies for which all the energies vanish looks just
like an integral that would occur in zero-temperature $3$-dimensional
QCD, and simple power counting at each ${\bf k} = 0$ reveals the
problem.  However, it seems that the problem goes away if one sums
over all thermal graphs.  Give each gluon a mass $m$.  Then, apart from
possible difficulties with doing this in a gauge theory, 
in the real-time formalism one can
derive the formula\defref\horgan{
I T Drummond, R R Horgan, P V Landshoff and A Rebhan, Physics Letters 
B398 (1997) 326}

$$
P = -\int_0^{\infty} dm^2 \int {d^4 q\over (2\pi)^4} { \epsilon (q_0)
\over e^{\beta q^0}-1}\;{\rm Im}
{1\over q^2 - m^2 - \hat\Pi (q, T, m)} \eqno (33)
$$

where $\hat\Pi$ is the self
energy of the gluon defined as in (24b) and $\epsilon (q_0)=\pm 1$
according to whether $q_0$ is positive or negative.  
As any divergence of $\hat\Pi$ now
appears in the denominator, the summation has made it harmless.  One
has to worry about the denominator possibly vanishing when $q=0$ and
$m=0$, but this will be rendered harmless by the $q^3$ appearing in $d^4q = q^3
dq d\Omega$.
\bigskip
{\bf Linear response theory}

Suppose that the thermal equilibrium of a plasma is disturbed by the
switching on at $t=0$ of an external electrostatic potential
$A^0_{{\rm ext}} (x)$.  Then the system's Hamiltonian acquires an extra
term

$$
H' (t) = \theta (t)\int d^3 x\, J_0 (x) A^0_{\rm ext} (x) \eqno (34)
$$

where $J_0$ is the charge density.  This will cause $J_0$ to change.
As it is a Heisenberg-picture operator, its equation of motion is

$$
{\partial J_0 \over \partial t} = i \left [ H + H', J_0 \right ] \eqno (35a)
$$

where $H$ is the original Hamiltonian.  When we take the thermal
average of this equation, the contribution from $H$ will disappear
because originally there was thermal equilibrium.  So the integrated
change in $\langle J_0 (x) \rangle$ at very large time is

$$
\delta \langle J_0 (x) \rangle = \int d^4 x' \,G_{R} (x-x')
A^0_{{\rm ext}} (x') \eqno (35b)
$$

where 

$$
G_R (x-x') = \theta (t-t') \langle [ J_0 (x), J_0 (x') ] \rangle \eqno
(35c)
$$

Taking the Fourier transform,

$$
\delta \langle J_0 (k) \rangle = G_R (k) \,A^0_{\rm ext} (k) \eqno (35d)
$$

We may express the retarded Green's function $G_R$ in terms of 
elements of the matrix Green's function ${\bf G}^{\mu\nu}$

$$
G_R = {\half} (G_{11}^{00} - G_{22}^{00} + G_{21}^{00} - G_{12}^{00}) 
\eqno (36)
$$

Each of the terms here may be calculated from pertubation theory.
However, it is simpler to express $G_R$ in terms of the 
function $\hat G^{00}(k)$ that appears in the diagonal matrix associated
with ${\bf G}^{00}$ (see (23)):

$$
G_R = n(k^0) e^{\half\beta |k^0 |} \{ 
(\hat G^{00}+\hat G^{00*})\sinh \half\beta |k^0 | + (\hat G^{00}-\hat G^{00*})\sinh \half\beta k^0 \} \eqno (37a)
$$

On the other hand

$$
G_{11} = n(k^0)e^{\half \beta |k^0 |} \{ 
 (\hat G^{00}+\hat G^{00*})\sinh\half\beta |k^0 |
+ (\hat G^{00}-\hat G^{00*})\cosh\half\beta k^0  \}
$$

So it is sufficient to calculate $G_{11}^{00}$ and change its imaginary
part by multiplying it by $\tanh {1\over 2}\beta k^0$.

\bigskip
{\sl I am grateful to Dr Arthur Hebecker for his help.}
\bigskip\goodbreak
\medskip\immediate\closeout\rfile\writestoppt
\baselineskip=10pt{{\bf References}}\bigskip{\frenchspacing%
\parindent=20pt\escapechar=` \input refs.tmp\bigskip}\nonfrenchspacing
\bye